\documentstyle[prd,aps,floats,epsfig,preprint]{revtex}
\tightenlines

\def\appendix{\par
 \setcounter{section}{0}
 \setcounter{subsection}{0}
 \def\thesection{Appendix \Alph{section}}
 \def\thesubsection{\Alph{section}.\arabic{subsection}}
 \def\theequation{\Alph{section}.\arabic{equation}}
 \setcounter{equation}{0}}

\begin{document}

\renewcommand{\thefootnote}{\fnsymbol{footnote}}

\title{Heavy Majorana neutrino production at electron-muon colliders}

\begin{flushright}
hep-ph/9906253 (revised version)\\
Phys. Lett. B {\bf 461} (1999) 248, {\em ibid} {\bf 471} (2000) 471E
\end{flushright}

\vspace{1.5cm}

\centerline{{\large \bf
Heavy Majorana neutrino production at electron-muon colliders}}

\vspace{0.8cm}

\centerline{ G.~Cveti\v c\footnote[1]{e-mail:
cvetic@physik.uni-bielefeld.de}}

\centerline{{\small \it 
Dept.~of Physics, Universit\"at Bielefeld,
33501 Bielefeld, Germany;}}
\centerline{{\small \it  
Dept.~of Physics, Universit\"at Dortmund,
44221 Dortmund, Germany}}

\vspace{0.5cm}

\centerline{C.S.~Kim\footnote[2]{e-mail:
kim@cskim.yonsei.ac.kr; http://phya.yonsei.ac.kr/\~{}cskim/}}

\centerline{{\small \it 
Dept.~of Physics, Yonsei University, 
Seoul 120-749, Korea}} 

\renewcommand{\thefootnote}{\arabic{footnote}}

\begin{abstract}
Possibilities for detecting heavy Majorana neutrinos ($N$'s)
at future $e{\mu}$ colliders are investigated.
In contrast to the $e^-e^+$ colliders (LEP200 and NLC),
the center-of-mass (CMS) energies achieved at $e{\mu}$ colliders 
can be much higher and the $Z$-mediated $s$ channel is excluded
automatically. This opens the attractive possibility of having
high production cross sections for $N$'s
and at the same time probing only the strength of charged current
couplings of $N$'s ($NWe$ and $NW{\mu}$).
The production cross sections and the expected numbers
of events for the reaction $e^{\mp} {\mu}^{\pm} \to NN \to 
W^{\pm} {\ell}^{\mp} W^{\pm} {\ell}^{\prime \mp}$
are calculated for various masses $M$ of the Majorana neutrinos and for
the CMS energies $\sqrt{s}\!=\!0.5$-$6.0$ TeV.
The values of the charged current coupling parameters are set equal to
their present upper bounds. We obtain reasonably high production cross
sections. Further, the effects of the off-shell intermediate $N$'s
turn out to be significant only at 
$\sqrt{s} \stackrel{>}{\approx} 2$ TeV.\\ 
PACS number(s): 14.60.St, 11.30.Fs, 13.10.+q, 13.35.Hb

\end{abstract}

\setcounter{equation}{0}
\newpage

\section{Introduction}

One of the basic questions in high energy physics is:
Are neutrinos Dirac or Majorana particles? 
In the absence of right-handed currents,
it is virtually impossible to
discern the nature of the light neutrinos \cite{Kayser}. 
However, if heavy neutrinos ($M\!\stackrel{>}{\sim}\!10^2$ GeV)
exist, then present and, even more so, future experiments
could establish whether such neutrinos are Majorana or Dirac. 
Many theoretical investigations have been
carried out \cite{Aguilaetal} into the
production cross sections of heavy Majorana neutrinos ($N$'s)
as predicted by various models, via $e^-e^+$, $e p$ and
$p p$ collisions,
and mostly within specific models or classes of models.
As a rule, it has been assumed that the center-of-mass (CMS) 
energy $\sqrt{s}$ in the process is high enough for the production 
of {\em on-shell\/} (OS) heavy $N$'s ($e^+e^-, ep, pp \to NN$). 
Further, the effects of the {\em off-shell\/} (nOS)
$N$'s have been ignored. 
We note that the detection of $NN$ pair production events
can be realized by identifying the final decay products (on-shell)
of the Majorana neutrinos -- primarily the leptons ${\ell}_j$'s
and the $W$'s. Thus, the reactions
${\ell}^{-}{\ell}^{\prime +} \to NN \to 
W^{\pm} {\ell}^{\mp}_i  W^{\pm} {\ell}^{\mp}_j
\to {\rm jets} + {\ell}^{\mp}_i {\ell}^{\mp}_j$, 
which violate the total lepton number,
would be a clear signal of the Majorana character of
the intermediate neutrinos (the latter need not be on-shell).
Here, ${\ell}$ and ${\ell}^{\prime}$ are the two
initial leptons, either $e^-e^+$ or $e{\mu}$.

In a previous paper \cite{CKK}, we demonstrated the
importance of taking into account the contributions
of {\em off-shell\/} intermediate $N$'s in such processes,
in $e^-e^+$ colliders such as LEP200 and future
Linear Colliders. Several advantages of $e{\mu}$ colliders
were pointed out by the authors of \cite{emu}.
In the present paper, we present some results of calculations
for the afore-mentioned reactions for $e{\mu}$
colliders. The main advantages of the $e{\mu}$ colliders,
in contrast to the $e^-e^+$ colliders, 
from our perspective, are the following: 
\begin{itemize}
\item
${\mu}$'s can easily be
accelerated to very high energies due to reduced synchrotron
radiation loss (since $m_{\mu}\!\gg\!m_e$);
\item
the $s$ channel ($Z$-mediated) is not present now.
\end{itemize}
Due to the first point, the event rates can be
increased because of the increased center-of-mass (CMS)
energy $\sqrt{s}$ of the process. Due to the second point,
the number of parameters on which the production rate
depends is reduced since only the couplings
of $N$'s to charged currents contribute (${\mu}WN$
and $eWN$ couplings).
The second point does not apply in the case of ${\mu}^-{\mu}^+$
colliders, the case which we therefore do not consider
in the present paper.

As in the previous paper \cite{CKK}, our analysis is rather
general, in the sense that we do not restrict ourselves
to any specific (classes of) models. Further, we use
basically the same set of programs developed by us there,
this time adjusted to the case of $e{\mu}$ colliders.
We stress that, in contrast to
the available literature, our numerical calculations 
include the effects of {\em off-shell\/} (nOS) intermediate $N$'s 
on the cross sections.\footnote{
This was the case also in our previous paper \cite{CKK}.}
This enables us to investigate
deviations from the previously known cross sections 
(with two on-shell intermediate $N$'s), in the ``2OS'' kinematic 
region ($\sqrt{s}\!>\!2M\!>\!2M_W$) where
both intermediate $N$'s can, but need not, be on shell (OS) 
-- these deviations are often referred to
as ``finite width effects''.
Further, our general
expressions allow us to calculate the cross sections
also in the ``1OS'' kinematic region
($2M\!>\!\sqrt{s}\!>\!M\!+\!M_W$) where at most one intermediate
$N$ can (but need not) be on shell, and even in the ``nOS'' region
($M\!+\!M_W\!>\!\sqrt{s}$) where both intermediate $N$'s
always have to be off shell (nOS).
For various choices of physical and kinematic parameters,
we calculate the total cross section for the mentioned
lepton-number violating reactions and the corresponding expected
numbers of events for $e{\mu}$ colliders. 

\section{Reaction amplitude and its square} 

\begin{figure}[htb]
\setlength{\unitlength}{1.cm}
\begin{center}
\epsfig{file=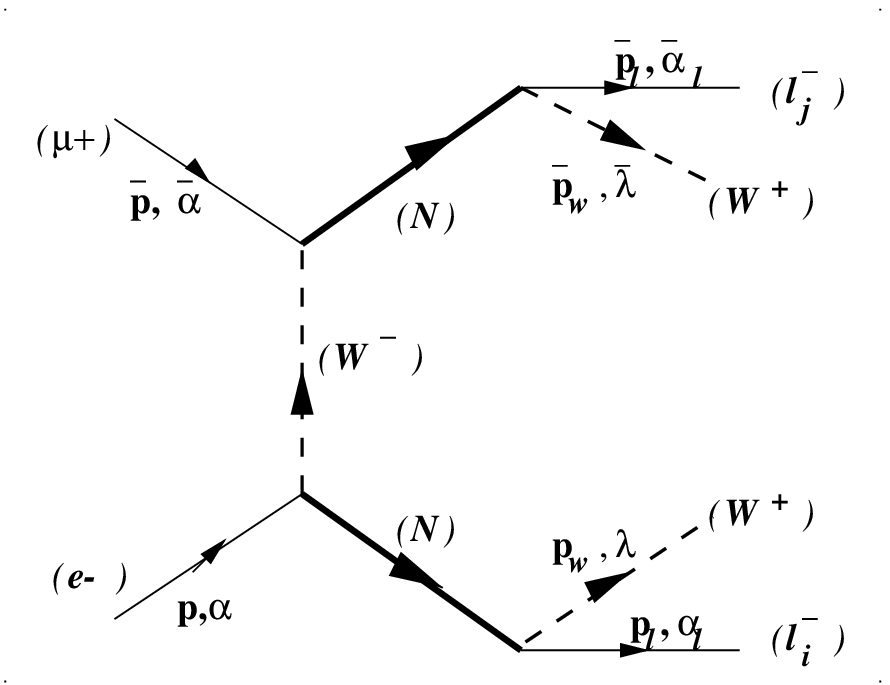, width=10.cm}
\end{center}
\vspace{-0.0cm}
\caption{\footnotesize 
Reaction 
$e^-(p,{\alpha}) {\mu}^+({\overline p},{\overline \alpha}) \to$ 
$N^{\ast} N^{\ast} \to W^+(p_w,{\lambda})
W^+({\overline p}_w,{\overline \lambda}) 
{\ell}_i^-(p_{\ell},{\alpha}_{\ell})
{\ell}_j^-({\overline p}_{\ell},{\overline \alpha}_{\ell})$. 
Attaching the legs of the four final particles to the two
$N$-propagators differently leads 
to the other contributing ($tu$-channel) 
diagrams. The CMS squared energy of the process is 
$s\!=\!(p\!+\!{\overline p})^2$.}
\label{Figemu1}
\end{figure}

Our starting point
is the rather general Lagrangian density for the
couplings of the heavy Majorana neutrinos $N$ with 
$W$'s and light leptons ${\ell}_j$
($j\!=\!1,2,3;$ ${\ell}_1^-\!=\!e^-$, 
${\ell}_2^-\!=\!\mu^-$, ${\ell}_3\!=\!{\tau}^-$):
\begin{equation}
{\cal {L}}_{N{\ell}W}(x) = - \frac{ g }{ 2 \sqrt{2} } \sum_{j=1}^3
B_L^{(j)} {\overline \ell}_j(x) 
{\gamma}^{\mu} ( 1 - {\gamma}_5 ) 
N(x) W_{\mu}^{(-)} + {\rm h.c.} \ ,
\label{NellW}
\end{equation}
where $B_L^{(j)}$'s can be regarded, at first, 
as basically free parameters (later we discuss some
possible constraints on $B_L^{(j)}$'s). In Eq.~(\ref{NellW}),
$g$ is the standard $SU(2)_L$ gauge coupling
parameter. We mention that the
$N{\ell}W$ couplings could include, in principle,
the right-handed parts, but this possibility will not be
considered here. Our choice would thus appear to suggest that
the considered $N$ (mass eigenstate) is primarily 
sequential, i.e., with the standard $SU(2)_L\!\times\!U(1)_Y$
assignments. However, our choice can approximately
describe also many other scenarios, including the one
in which $N$ has an admixture of $SU(2)_L$-singlet component,
provided the right-handed coupling analogous to (\ref{NellW})
is reasonably suppressed. Further, it turns out that
parameters $B_L^{(j)}$'s affect the final results only
via the combinations
\begin{equation}
H^{\prime} \equiv |B_L^{(1){\ast}} B_L^{(2)}| \ , \quad
H \equiv \sum_{j=1}^3 |B_L^{(j)}|^2 \ .
\label{HpH}
\end{equation}
We will restrict ourselves to considering the reactions
$e^-{\mu}^+ \to NN \to W^{\pm} {\ell}_i^{\mp}W^{\pm} {\ell}_j^{\mp}
(\to {\rm jets} + {\ell}_i^{\mp} {\ell}_j^{\mp})$, which
represent an explicit violation of the total lepton
number -- characteristic for the intermediate heavy
{\em Majorana\/} neutrinos. The reaction involves 
only the $t\!+\!u$ (shortly: $tu$) channel --
cf.~Fig.~\ref{Figemu1}. The $s$ channel ($Z$-mediated) is not allowed.

For the calculation of the invariant amplitude
${\cal {M}}_{fi}$ (shortly: ${\cal {M}}$)
for the $tu$ channel,\footnote{ 
${\cal M}_{fi}$ is defined by \cite{Peskinetal}:  
$\langle fin | S | in \rangle \equiv  {\rm i} (2 \pi )^4
\delta^{(4)}(P_{in} - P_{fin}) {\cal {M}}_{fi}$,
for $|fin \rangle\!\not=\!|in \rangle$.}
we used the 4-component
spinors $u^{(\alpha)}(q)\!\equiv\!u(q \alpha)$ 
and $v^{(\alpha)}(q)\!\equiv\!v(q \alpha)$ as
defined by Itzykson and Zuber \cite{IZ} (I.Z.),
but with the normalization convention as given
in \cite{Peskinetal}, i.e.,
$u\!=\!\sqrt{2 m_f} u_{\rm I.Z.}$ and 
$v\!=\!\sqrt{2 m_f} v_{\rm I.Z.}$, where $m_f$ is the
mass of the fermion. Further, for
the spin component quantum numbers of spinors, we use the 
following notational convention for tildes:
${\widetilde{\alpha}}\!=\!1, 2 \Leftrightarrow \alpha\!=\!2, 1$.
The $tu$-channel amplitude ${\cal M}^{(tu)}$ is
\begin{eqnarray}
\lefteqn{
{\rm i} {\cal M}^{(tu)} = \frac{4 M A^{(tu)} }
{ [ (p\!-\!p_{\ell}\!-\!p_w)^2\!-\!M_W^2\!+\!{\rm i} {\Gamma}_W M_W ] }
(-1)^{{\overline \alpha}_{\ell}} P_N(p_{\ell}p_w) 
P_N({\overline p}_{\ell}{\overline p}_w) \times }
\nonumber\\
&&{\Big \lbrace} \left[ {\overline u}(p_{\ell};{\alpha}_{\ell})
( {\varepsilon \llap /} 
{p \llap /}_w\!+\!2 p_{\ell}\!\cdot\!{\varepsilon} ) 
{\gamma}^{\nu} (1\!-\!{\gamma}_5) u(p;{\alpha}) \right] 
\left[ {\overline v}({\overline p};{\overline \alpha})
{\gamma}_{\nu} (1\!+\!{\gamma}_5) 
{{\overline \varepsilon} \llap /} (1\!+\!{\gamma}_5) 
v({\overline p}_{\ell};{\widetilde {\overline \alpha}}_{\ell}) \right]
\nonumber\\
&&+ \frac{ (p_{\ell}\!+\!p_w)^2 }{M_W^2} 
\left[{\overline u}(p_{\ell};{\alpha}_{\ell}) {\varepsilon \llap /}
(1\!-\!{\gamma}_5) u(p;{\alpha}) \right]
\left[{\overline v}({\overline p};{\overline \alpha}) 
( {{\overline p} \llap /}_w {{\overline \varepsilon} \llap /}\!+\!2 
{\overline p}_{\ell}\!\cdot\!{\overline \varepsilon} )
(1\!+\!{\gamma}_5)
v({\overline p}_{\ell};{\widetilde {\overline \alpha}}_{\ell}) \right]
{\Big \rbrace}\!+\!\ldots
\label{tuAmpl1}
\end{eqnarray}
Here we use the notations of Fig.~\ref{Figemu1}.
Further, ${q \llap /}\!\equiv\!q_{\nu} {\gamma}^{\nu}$;
$\varepsilon_{\nu}\!\equiv\!{\varepsilon}_{\nu}^{(\lambda)}(p_w)$
and ${\overline \varepsilon}_{\nu}\!\equiv\!{\varepsilon}_
{\nu}^{(\overline \lambda)}({\overline p}_w)$ are the real
polarization vectors \cite{IZ} of the final $W$'s, with polarizations 
$\lambda,{\overline \lambda}\!=\!1,2,3$. 
$M$ is the mass of $N$'s.
Parameter $A^{(tu)}$ in Eq.~(\ref{tuAmpl1}) contains the strengths
of the couplings of the two Majorana neutrinos in the $tu$ channel
\begin{equation}
A^{(tu)} = \frac{g^4}{64} B_L^{(1)\ast} B_L^{(2)} B_L^{(i)}B_L^{(j)} 
{\rm i} \lambda_M \ ,
\label{Atu}
\end{equation}
where ${\lambda}_M$ is the phase factor in the
Fourier decomposition of the Majorana field $N(x)$
(cf.~\cite{Kayseretal}; $|\lambda_M|^2\!=\!1$). $P_N$
in Eq.~(\ref{tuAmpl1}) is the (scalar) denominator of the
propagator of $N$ 
\begin{equation}
P_N(p_{\ell}p_w) =  
1/[ (p_{\ell}\!+\!p_w)^2\!-\!M^2\!+\!{\rm i} M {\Gamma}_N ] \ ,
\label{PN}
\end{equation}
where ${\Gamma}_N$ is the total decay width of $N$.
The dots in Eq.~(\ref{tuAmpl1}) 
at the end stand for three analogous terms,
obtained from the expression explicitly written above by replacements:
(I) $(p_w,{\varepsilon})\!\leftrightarrow\!({\overline p}_w,
{\overline \varepsilon})$; 
(II) $(p_{\ell},\alpha)\!\leftrightarrow\!({\overline p}_{\ell},
{\overline \alpha})$ and overall factor $(-1)$;
(III) combined replacements (I) and (II). In expression
(\ref{tuAmpl1}) we neglected the masses of all the charged
leptons appearing ($e, \mu, \tau$).

We can reexpress any of the above terms
in ${\cal M}^{(tu)}$ in alternative forms, by applying
the following general identities:
\begin{equation}
- {\rm i} {\gamma}^2 u(q {\alpha})^{\ast}\!=\!(-1)^{\alpha} 
v(q {\widetilde {\alpha}}), \
- {\rm i} {\gamma}^2 v(q {\alpha})^{\ast}\!=\!(-1)^{\widetilde {\alpha}}
u(q {\widetilde {\alpha}}) ,
\label{CPtransf}
\end{equation}
and $({a \llap /} {b \llap /} )^T\!=\!- {\gamma}^0 {\gamma}^2
({b \llap /}{a \llap /}) {\gamma}^2 {\gamma}^0$,
where the Dirac basis and the conventions of \cite{IZ}
are used for ${\gamma}^{\mu}$'s.
We use (\ref{CPtransf}), for example, if we want to employ,
in scalar expressions in square brackets of Eq.~(\ref{tuAmpl1}), 
$u({\overline p};{\widetilde {\overline \alpha}})$ and
${\overline u}({\overline p}_{\ell};{\overline \alpha}_{\ell})$ 
instead of ${\overline v}({\overline p};{\overline \alpha})$ and
$v({\overline p}_{\ell};{\widetilde {\overline \alpha}})$. 
Such transformations are very convenient when calculating 
$\langle |{\cal M}|^2 \rangle\!=\!\langle |{\cal M}^{(tu)}|^2 \rangle$.
Using the afore-mentioned tranformations judiciously,
we can always end up with traces of $4 \times 4$ matrices
involving $u(q,{\beta}) {\overline u}(q,{\beta})\!=\!{q \llap /}$
and/or $v(q,{\beta}) {\overline v}(q,{\beta})\!=\!{q \llap /}$
($q\!=\!p,{\overline p},\ldots$; 
${\beta}\!=\!{\alpha},{\overline \alpha},\ldots\!=\!1,2$).
We performed these traces by writing programs in
{\em Mathematica\/} \cite{Mathematica} and employing
there subroutine {\em Tracer\/} \cite{Tracer}.
Single traces involve up to fourteen ${\gamma}^{\mu}$ matrices
and $(1\!-\!{\gamma}_5)$ matrix. 

The expressions for $\langle |{\cal M}|^2 \rangle$,
obtained in the way described above, 
were then rewritten with short notations for various
scalar products and for the contracted Levi-Civita tensors.
They were then translated into an optimized FORTRAN form
by employing {\em Maple\/} \cite{Maple}, and fed into our main
program for calculation of the total cross sections ${\sigma}$.
The integrand $\langle |{\cal M}|^2 \rangle$ is a long expression,
extending over tens of pages when printed out.
The main program uses subroutines {\em rambo\/} \cite{rambo}
and {\em vegas \/} \cite{vegas} for numerical integration.

\section{Numerical calculations and results}

We stress that our general program can now deal with all
kinematic situations:\\
1.) those where both intermediate $N$'s can (but need not) be on shell 
[2OS kinematic region: $\sqrt{s}\!>\!2 M\!>\!2 M_W$];\\
2.) those where only at most one $N$ can (but need not) be on shell
[1OS kinematic region: $2 M\!>\!\sqrt{s}\!>\!M\!+\!M_W$];\\ 
3.) those where both $N$'s
must be off shell [nOS kinematic region: $M\!+\!M_W\!>\!\sqrt{s}$]. 

However, by the same methods as described above, we also
calculated the amplitudes ${\cal M}$ and their averaged 
squares $\langle |{\cal {M}}|^2 \rangle$
when one $N$, or both $N$'s, are explicitly put on shell 
(1OS, 2OS expressions, respectively). These expressions 
turn out to be much shorter than the general (nOS)
expressions with both $N$'s propagating. Analogously as 
for the nOS expressions (cf. previous Section), also the
1OS and 2OS expressions for $\langle |{\cal {M}}|^2 \rangle$
were calculated with help of the subroutine {\em Tracer\/}
and were then fed into separate FORTRAN programs where
they were combined with the {\em rambo\/} and {\em vegas\/}
subroutines. These (1OS, 2OS) programs allow us to calculate
the sum of cross sections for the considered reactions
$e^-{\mu}^+ \to NN \to W^+ {\ell}_i^- W^+ {\ell}_j^-$ 
($i,j\!=\!1,2,3$)
where now one, or both, intermediate $N$'s are explicitly 
on shell. In fact, the 1OS expression for 
$\langle |{\cal {M}}|^2 \rangle$, for the sum of reactions
$e^-{\mu}^+ \to N N(OS) \to W^{+} {\ell}_i^- N(OS)$
($i\!=\!1,2,3$),
is multiplied simply by the branching ratio $Br$
for the sum of the decay modes
$N(OS) \to W^+ {\ell}_j^-$ ($j\!=\!1,2,3$);
the 2OS expression for the reaction
$e^-{\mu}^+ \to N(OS) N(OS)$ is multiplied by $(Br)^2$.

For numerical calculations, the input
were values of $\sqrt{s}$, $M$,
and of the ``mixing'' parameters
$H^{\prime}$ and $H$  [cf.~Eqs.~(\ref{NellW})-(\ref{HpH})].
Parameter $H^{\prime}$ is a combined measure of the strengths of the
$eWN$ and ${\mu}WN$ couplings and it affects the ($tu$) amplitude
crucially ($\propto H^{\prime}$). Parameter $H$
affects the total $\langle |{\cal M}|^2 \rangle$
which is then {\em formally\/} proportional to $H^2$ 
(once $H^{\prime}$ is fixed). This is so because we sum over all the
flavors of the two final light leptons ${\ell}_i^-$,
${\ell}_j^-$. All in all, in $\langle |{\cal M}|^2 \rangle$
we thus average over the spin components (${\alpha},
{\overline \alpha}$) of
the initial $e^-$ and ${\mu}^+$ (factor 1/4),
sum over the polarizations (${\lambda}, {\overline \lambda}$)
of the two final $W^+$'s and over the spin components 
(${\alpha}_{\ell}, {\overline \alpha}_{\ell}$)
{\em and\/} the flavors ($i,j\!=\!1,2,3$) of 
the two final light leptons.
In the general (nOS) expression,
we have to include an additional factor $1/2$
to avoid double-counting of the two $W^+$'s,
and factor $1/2$ to avoid double-counting of
the final leptons (i.e., of ${\ell}_i^- {\ell}_i^-$,
or of twice ${\ell}_i^- {\ell}_j^-$ when $i\!\not=\!j$).
The $\langle |{\cal M}|^2 \rangle$ thus defined
can then be integrated over the entire phase
space of the final particles (massive $W$'s
and massless ${\ell}$), 
and there is no double-counting\footnote{
In the simpler 1OS and 2OS expressions for
$\langle |{\cal M}|^2 \rangle$, these factors are different.
For the 2OS expression, the factor is $(1/2)$ because the
two intermediate {\em on-shell\/} $N$'s are indistinguishable,
and thus their decay product pairs ($W^+{\ell}_i^-$, $W^+{\ell}_j^-$) 
are also kinematically indistinguishable. For the 1OS expression, 
in the 1OS kinematic region, all four final particles are 
distinguishable via kinematics (i.e., invariant masses)
of $W^+{\ell}^-$ pairs, and the factor is $1$; 
in the 2OS kinematic region, the two decay product pairs 
($W^+{\ell}_i^-$, $W^+{\ell}_j$)
become kinematically indistinguishable, and the factor
is $(1/2)$.}
in the obtained total ${\sigma}$. 
 
Further, the total decay width ${\Gamma}_N$ of $N$'s, 
appearing in the denominators of their propagators,
was determined by us at the tree level and under the assumption that
the only (dominant) decay modes are $N \to W^{\pm} {\ell}_j^{\mp}$
($j\!=\!1,2,3$). Incidentally, this means that the
branching ratio for the sum of the decay modes
$N\!\to\!W^+ {\ell}_j^-$ ($j\!=\!1,2,3$) is
simply $Br\!=\!1/2$, the value that is then also used in the
1OS and 2OS programs for cross sections (cf.~discussion above).
In this framework, we have ${\Gamma}_N\!\propto\!H$, i.e.,
$H$ is the decay width parameter of the $N$'s.

\begin{figure}[htb]
\setlength{\unitlength}{1.cm}
\begin{center}
\epsfig{file=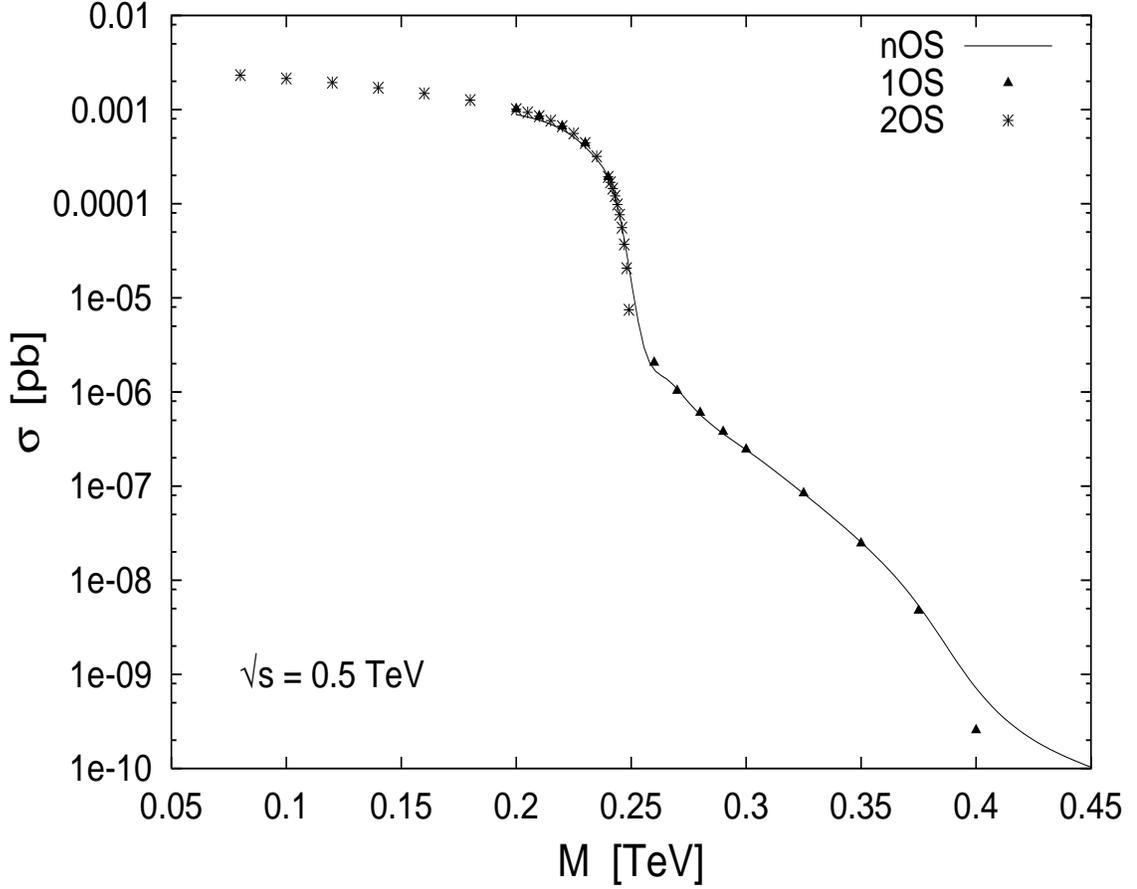,height=12.5cm,width=15.5cm}
\end{center}
\vspace{-0.4cm}
\caption{\footnotesize 
Sum of the total cross sections for the reactions 
$e^-{\mu}^+ \to N N \to W^+ W^+ {\ell}_i^- {\ell}_j^-$
($i,j\!=\!1,2,3$; ${\ell}_1\!=\!e$, ${\ell}_2\!=\!{\mu}$, 
${\ell}_3\!=\!{\tau}$), as function of the Majorana neutrino mass $M$,
for the CMS process energy $\sqrt{s}\!=\!0.5$ TeV.
Summation over the final, and average over the initial 
polarizations were carried out. The values of the
``mixing'' parameters $H^{\prime}$ and $H$ of Eq.~(\ref{HpH})
were set equal to the upper bounds (\ref{HpHrestr}).
The results of the general (nOS) program are displayed
as full lines, in various kinematic regions.
The results of the simpler 1OS program are displayed
as triangles, in the 1OS and partly in the 2OS kinematic
regions. The results of the short 2OS program are
displayed as crosses, in the 2OS kinematic region.}
\label{Figemu2}
\end{figure}

\begin{figure}[htb]
\setlength{\unitlength}{1.cm}
\begin{center}
\epsfig{file=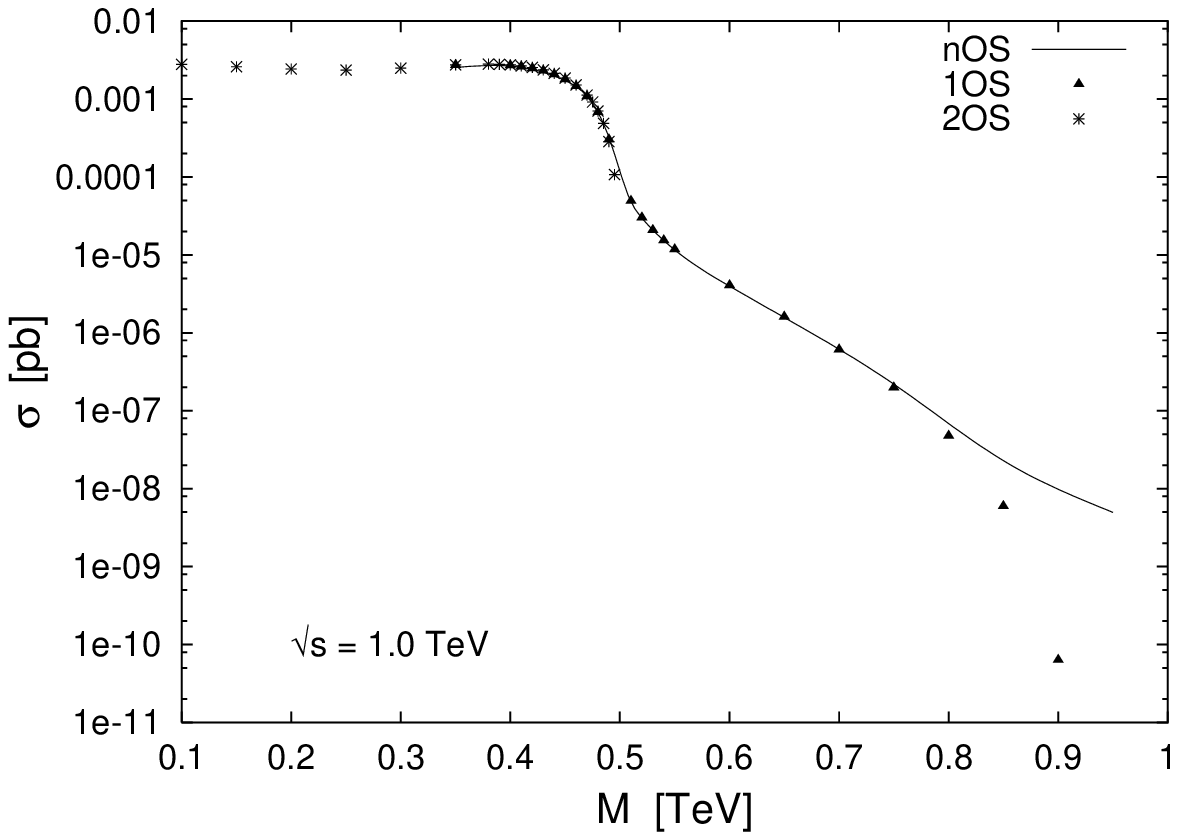,height=12.5cm,width=15.5cm}
\end{center}
\vspace{-0.4cm}
\caption{\footnotesize
Same as Fig.~\ref{Figemu2}, but for $\sqrt{s}\!=\!1.0$ TeV.}
\label{Figemu3}
\end{figure}

\begin{figure}[htb]
\setlength{\unitlength}{1.cm}
\begin{center}
\epsfig{file=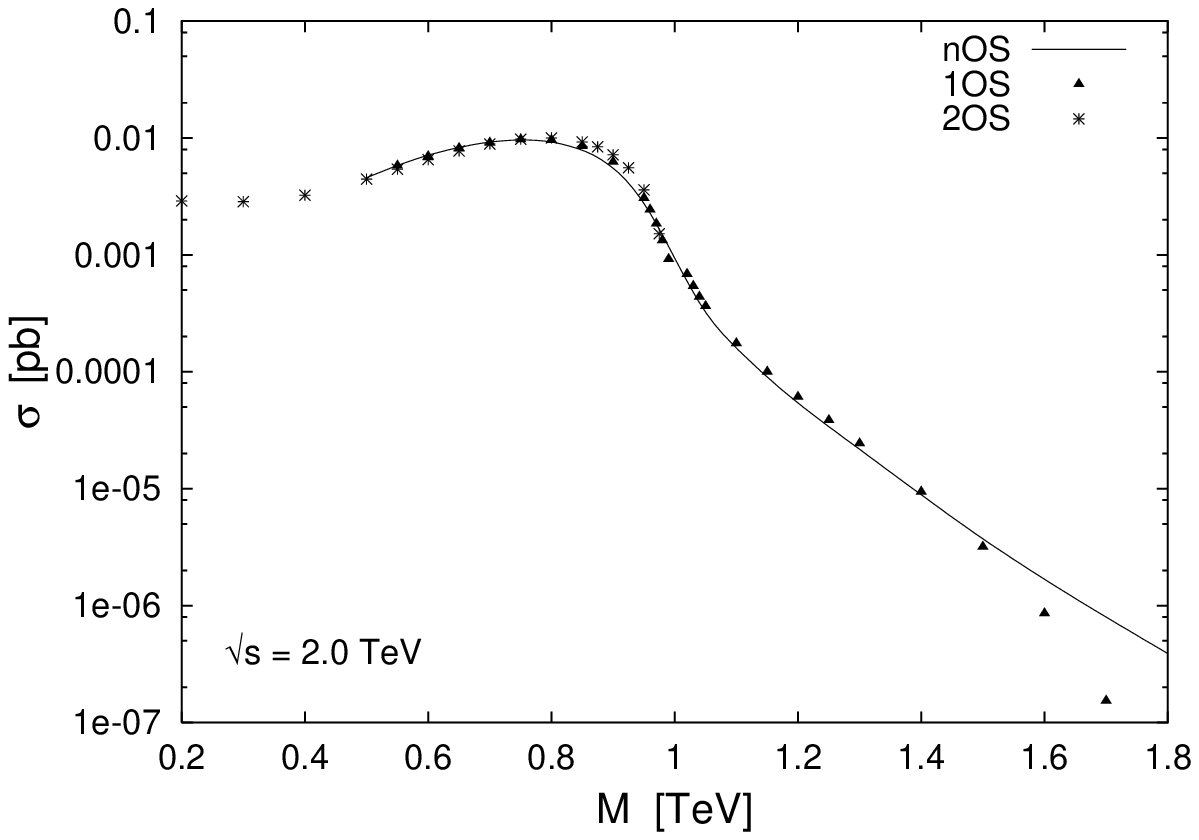,height=12.5cm,width=15.5cm}
\end{center}
\vspace{-0.4cm}
\caption{\footnotesize
Same as Fig.~\ref{Figemu2}, but for $\sqrt{s}\!=\!2.0$ TeV.}
\label{Figemu4}
\end{figure}

\begin{figure}[htb]
\setlength{\unitlength}{1.cm}
\begin{center}
\epsfig{file=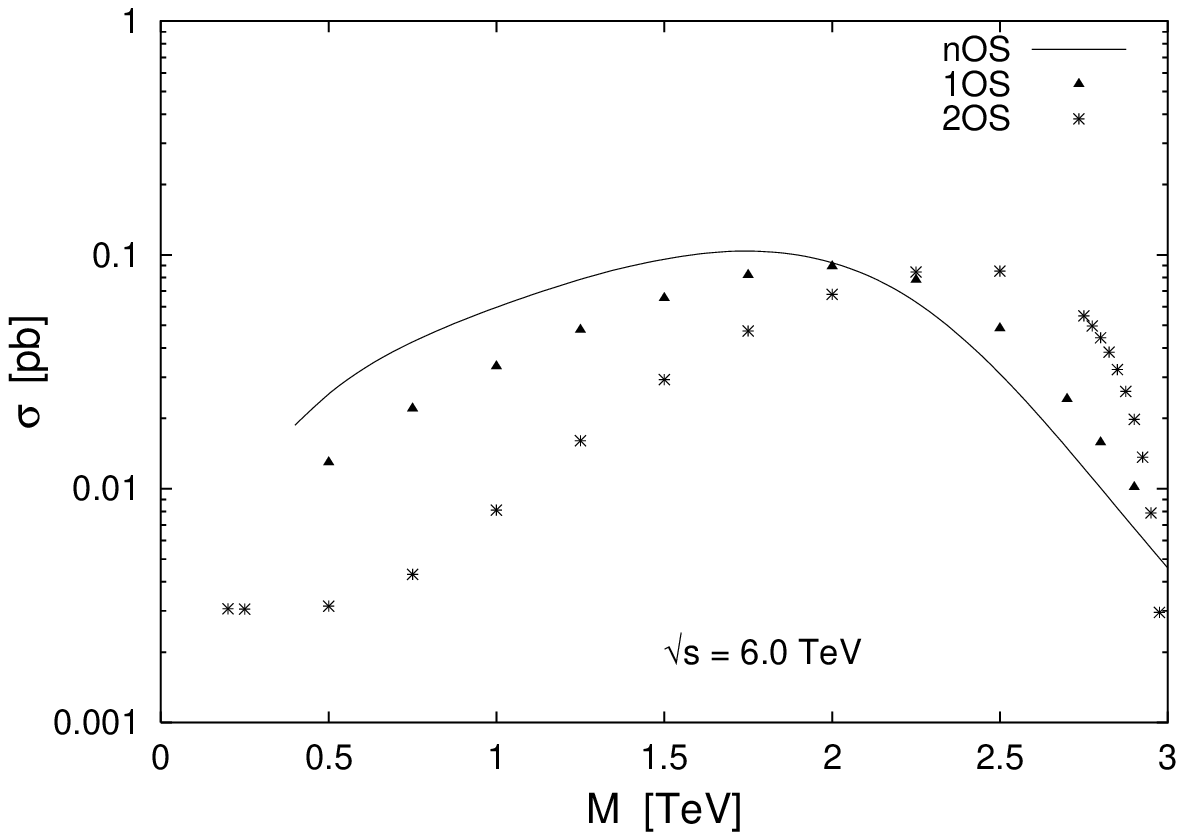,height=12.5cm,width=15.5cm}
\end{center}
\vspace{-0.4cm}
\caption{\footnotesize
Same as Fig.~\ref{Figemu2}, but for $\sqrt{s}\!=\!6.0$ TeV.}
\label{Figemu5}
\end{figure}

\begin{figure}[htb]
\setlength{\unitlength}{1.cm}
\begin{center}
\epsfig{file=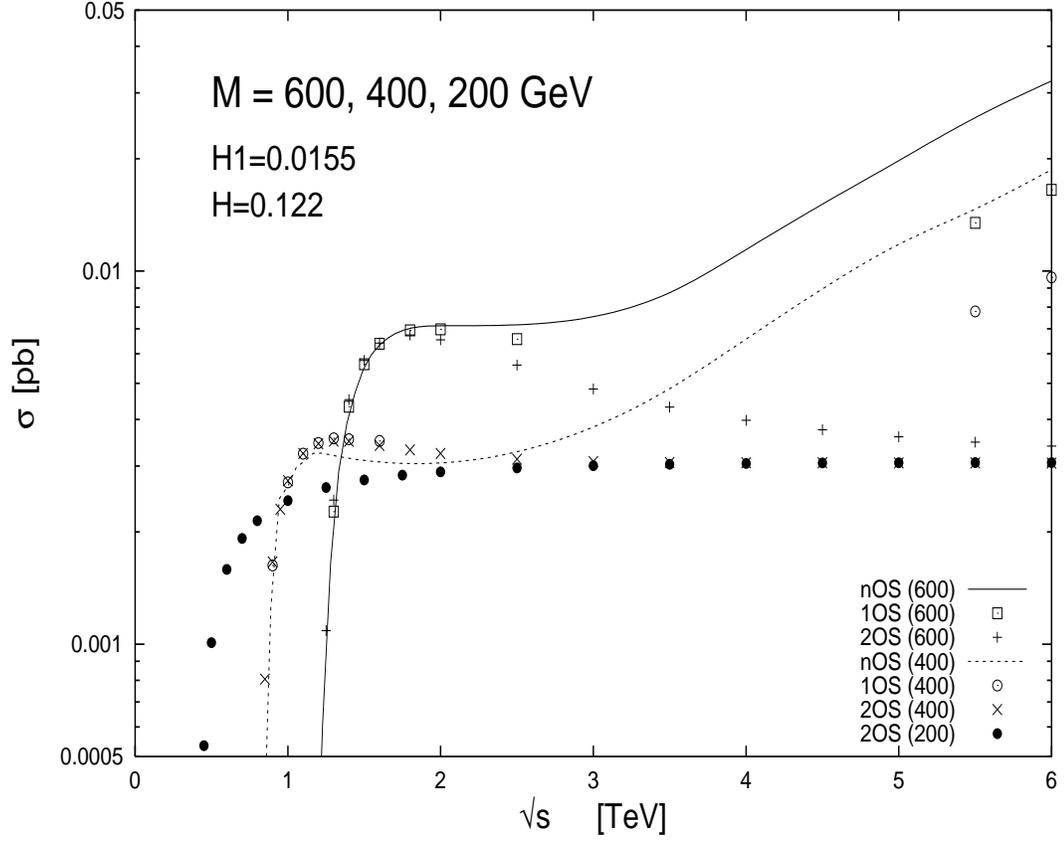,height=12.4cm,width=15.5cm}
\end{center}
\vspace{0.cm}
\caption{\footnotesize 
Sum of the mentioned total cross sections for the reactions 
$e^-{\mu}^+ \to N N \to W^+ W^+ {\ell}_i^- {\ell}_j^-$
($i,j\!=\!1,2,3$; ${\ell}_1\!=\!e$, ${\ell}_2\!=\!{\mu}$, 
${\ell}_3\!=\!{\tau}$), as function of 
the CMS process energy $\sqrt{s}$, for
Majorana neutrino mass $M\!=\!600$, $400$ and $200$ GeV.}
\label{Figemu6}
\end{figure}

In large classes of models, in which heavy neutrinos
are sequential or have exotic $SU(2)\!\times\!U(1)$
assignments, the values of the ``mixing'' parameters 
$H^{\prime}$ and $H$ [cf.~Eq.~(\ref{HpH})] are severely restricted by 
available experimental data (LEP and low-energy
data) \cite{LL,constr}:  
\begin{equation}
H^{\prime} \stackrel{<}{\approx} 0.0155 \ , \quad 
H \stackrel{<}{\approx} 0.122 \ .
\label{HpHrestr}
\end{equation}
For orientation, we used for the values of $H^{\prime}$ and $H$
these two upper bounds. The numerical results for
the cross sections ${\sigma}$ of the reactions
$e^- {\mu}^+ \to N N \to W^+ {\ell}_i^- W^+ {\ell}_j^-$
(summed over $i,j\!=\!1,2,3$) 
are depicted in Figs.~\ref{Figemu2}-\ref{Figemu6}. 

In Figs.~\ref{Figemu2}-\ref{Figemu5}, the cross sections ${\sigma}$
are presented as functions of the Majorana neutrino mass $M$, at five
different CMS reaction energies $\sqrt{s} = 0.5, 1, 2, 6$ TeV,
respectively.
Calculations were performed in various kinematic regions (nOS, 1OS, 2OS
regions) with the general (nOS) program which has the
propagating intermediate $N$'s with finite width 
[cf.~Eqs.~(\ref{tuAmpl1}), (\ref{PN})].
Further, in the 1OS and 2OS kinematic regions, the
cross sections were calculated also with the 1OS program
(in which one $N$ is explicitly on shell). In the 
2OS kinematic region, the cross sections were performed 
in addition with the 2OS program (in which both $N$'s are 
explicitly on shell). Comparing the results of the general 
nOS program with those of the
1OS and 2OS programs, we see from these Figures that
the contributions of the configurations with off-shell
Majorana neutrinos become important only at high 
$\sqrt{s} \geq 2$ TeV. For example, when $\sqrt{s}\!=\!2.0$ TeV,
the results of the nOS program in the 2OS kinematic region
($M\!<\!\sqrt{s}/2$)
differ from those of the 2OS (and 1OS) programs already by up to
$30 \%$. One of the reasons for this trend lies in the fact
that higher $\sqrt{s}$ (at a given $M$) mean that more off-shell
configurations are available for the intermediate $N$'s.
Further, in Figs.~\ref{Figemu2}-\ref{Figemu4}
we see the slope increase associated with the
change of kinematic regions: at
$M\!\approx\!\sqrt{s}/2$ (onset of the 2OS region).
In Fig.~\ref{Figemu5}, only the 2OS kinematic region
is included.

In connection with Figs.~\ref{Figemu2}-\ref{Figemu5},
one technical difficulty
with the general (nOS) program should be mentioned.
The results of the nOS program were not depicted in the lowest
parts of the 2OS kinematic regions 
($M$ much lower than $\sqrt{s}/2$).
In these regions of low $M$, the nOS
program requires exceedingly high statistics
to avoid large numerical uncertainties.
These numerical uncertainties
arise because, for low $M$'s ($\sim$$M_W$) 
only a very limited part of the rambo-generated phase
space of final particles contributes to the 
cross section (that would be true even when ${\Gamma}_N$
were, artificially by hand, prevented from approaching
the zero value when $M\!\to\!M_W$). An additional 
important reason for these uncertainties lies in the fact
that the integrand becomes very singularly peaked in
small subregions of that mentioned part of the phase space. 
This is so because ${\Gamma}_N$ becomes small for small $M$'s, 
and consequently the absolute square of expression (\ref{PN}) 
is then almost equal to
$[\pi/(M \Gamma_N)] \delta[(p_{\ell}\!+\!p_w)^2\!-\!M^2]$. 

In Fig.~\ref{Figemu6} we plot the 
cross sections as function of the CSM energy $\sqrt{s}$,
at fixed chosen values of the Majorana mass $M\!=\!200$,
$400$ and $600$ GeV. Here we see that,
at a given Majorana neutrino mass $M$, the full cross section
usually (but not always) increases with increasing $\sqrt{s}$. 
At high CSM energies $\sqrt{s}\!>\!2 M$ (2OS kinematic region), 
the cross section changes only slowly with increasing $\sqrt{s}$.
In Fig.~\ref{Figemu6}, for $M\!=\!200$ GeV only results of the 2OS
program (in which both $N$'s are put on shell)
are displayed, since the general (nOS) and the
1OS programs show up the afore-mentioned numerical instabilities 
and large numerical uncertainties at so low values of $M$.
Even for $M\!=\!400$ GeV, the nOS program has significant numerical 
uncertainties ($\sim$$10 \%$) for ${\sigma}$
in the region ${\sqrt{s}}\!=1$-$2$ TeV,
but these results were included nonetheless.

One may ask how the depicted results change once
we change the values of the ``mixing'' parameters $H^{\prime}$ and $H$
[cf.~Eqs.~(\ref{HpH}), (\ref{HpHrestr})].
It turns out that, for the 2OS kinematic region
($\sqrt{s}\!>\!2M\!>\!2M_W$), the cross
sections (and distributions) as given by the general (nOS) program
depend on the parameter $H$ only weakly. This can be
understood first in the formal way, because the absolute square
of each of the two intermediate $N$-propagators 
(when ${\Gamma}_N\!\ll\!M$)
is approximately proportional to ${\Gamma}_N^{-1}\!\propto\!H^{-1}$
(cf.~previous discussion), while the sum of the
squares of the amplitudes without 
$N$-propagators is proportional to $H^2$
-- the latter because of the two decay vertices 
$N \to W^+ {\ell}_j^-$ and $N \to W^+ {\ell}_i^-$
(cf.~Fig.~\ref{Figemu1}) which are squared and summed ($i,j\!=\!1,2,3$).
This approximate $H$-independence can be expected on physical grounds:
These cross sections are, for not too high $\sqrt{s}$, 
dominated by the configurations 
with two on-shell $N$'s; but the absolute squares of amplitudes
for $e^-{\mu}^+ \to N(OS) N(OS)$
[combined with the decay branching ratios $(1/2)^2$ for the $N(OS)$'s]
are exactly $H$-independent. Parameter $H$,
being responsible for the finite width
(note: ${\Gamma}_N \propto H$), is apparently one of the 
two principal parameters responsible in the 2OS kinematic region for the
deviation of the cross section 
from the pure 2OS cross section (the other is $\sqrt{s}$). 
On the other hand, 
in the 1OS kinematic region ($2 M\!>\!\sqrt{s}\!>\!M\!+\!M_W$),
the $H$-dependence of the cross section becomes quite strong
(approximately $\propto$$H$), and in the nOS region 
($M\!+\!M_W\!>\!\sqrt{s}$) even more so ($\propto$$H^2$).

On the other hand, the dependence of the cross sections
on the parameter $H^{\prime}$ [cf.~Eqs.~(\ref{HpH}), (\ref{Atu})]
is always strong ($\propto$$H^{{\prime} 2}$).

\section{Summary and conclusions}

We investigated the reactions 
$e^-{\mu}^+ \to N N \to W^+ {\ell}^- W^+ {\ell}^{\prime -}$
where the intermediate $N$'s are heavy Majorana neutrinos
(with mass $M\!\stackrel{>}{\sim}\!10^2$ GeV) 
and ${\ell}, {\ell}^{\prime}$ are light charged
leptons ($e, \mu, \tau$). These processes violate the
total lepton number and, if detected, 
would represent a clear signal of the
Majorana character of (heavy) neutrinos. 
In contrast to
the expressions available in the literature so far where
the intermediate $N$'s are taken to be {\em on shell\/}, we
calculated the amplitudes of these processes by employing
finite width propagators of the $N$'s. This enabled us
to calculate numerically the full cross section for the
reactions, i.e., including the effects of the 
{\em off-shell\/} intermediate $N$'s. In contrast
to the analogous reactions in $e^-e^+$ 
(and ${\mu}^-{\mu}^+$) colliders,
the considered reactions do not include a ($Z$-mediated)
$s$ channel, but only a ($W$-mediated) $tu$ channel,
thus allowing us to reduce the number of relevant
unknown coupling parameters.
If the CMS process energies $\sqrt{s}$ are large ($\geq 2$ TeV),
the effects of the off-shell intermediate $N$'s
can be significant even in the 2OS kinematic
region, i.e., in the region
$\sqrt{s}\!>2M$ ($>2 M_W$) where the configurations
with both $N$'s being on shell are allowed. 
Further, the number of such events at future
$e{\mu}$ colliders would be in general high enough
for detection.
For example, if assuming the
integrated luminosity $10^4 \ {\rm pb}^{-1}$,
and Majorana mass $M\!=\!200$ GeV ($400$ GeV),
we would get about $10$, $24$, $29$, $31$
($0$, $27$, $32$, $188$)
events at $\sqrt{s}\!=\!0.5, 1.0, 2.0$ and $6.0$ TeV,
respectively. These numbers have to be multiplied
by two if we include also the events with
two positively charged leptons in the final state
(the latter events also violate the total lepton number
conservation, of course).
We should stress that we used the highly suppressed
values of the ``mixing''
parameters $H^{\prime}\!=\!0.0155$ and $H\!=\!0.122$.
These values are the maximal values allowed in theories
where the heavy neutrinos are either sequential or
have exotic $SU(2)\!\times\!U(1)$ assignments.
If these bounds do not apply, the number of events
can increase dramatically since it is proportional
to $H^{\prime 2}$.

In the paper, we ignored the questions connected with the
experimental difficulties of detecting the discussed
process unambiguously. In particular, there are
problems connected with the identification of the
(on-shell) $W$'s and some of the light charged leptons (${\tau}$'s). 
Further, we ignored the possibility that the Majorana
neutrino mass $M$ is very low -- below $M_W$
(the $W^+$'s are then off shell) \cite{privcomm}.
In such interesting cases of low $M$, however, additional experimental 
problems would arise in the identification of the process
since the two $W^+$'s are now intermediate off-shell particles.

\vspace{0.7cm}

{\noindent{\bf Acknowledgments:}}\\
We would like to thank V.~Barger for his helpful comments.
G.C.~acknowledges support by the BMBF project 05-7DO93P7
during the initial stage of this work, and he
thanks Prof.~D.~Schildknecht
for the hospitality of Bielefeld University during the
completion of this work.
The work of C.S.K. was supported 
in part by KRF Non-Directed-Research-Fund, Project No.~1997-001-D00111,
in part by the BSRI Program, Ministry of Education, 
Project No.~98-015-D00061,
in part by the KOSEF-DFG large collaboration project, 
Project No.~96-0702-01-01-2.

\newpage

\end{document}